\documentclass[aps,pra,twocolumn,superscriptaddress,longbibliography,nofootinbib]{revtex4-2}
\usepackage{amsmath,amssymb,graphicx,color}

\begin{document}

\title{Multipartite reference-frame-independent quantum cryptographic communication}

\author{Donghwa Lee}
\email{fairytale095@gmail.com}
\affiliation{Center for Quantum Technology, Korea Institute of Science and Technology (KIST), Seoul 02792, Republic of Korea}

\author{Kyujin Shin}
\affiliation{Center for Quantum Technology, Korea Institute of Science and Technology (KIST), Seoul 02792, Republic of Korea}
\affiliation{Materials Research \& Engineering Center, R$\&$D Division, Hyundai Motor Company, Uiwang 16082, Republic of Korea}

\author{Hyang-Tag Lim}
\affiliation{Center for Quantum Technology, Korea Institute of Science and Technology (KIST), Seoul 02792, Republic of Korea}
\affiliation{Division of Quantum Information, KIST School, Korea University of Science and Technology, Seoul 02792, Republic of Korea}

\author{Yosep Kim}
\affiliation{Center for Quantum Technology, Korea Institute of Science and Technology (KIST), Seoul 02792, Republic of Korea}
\affiliation{Department of Physics, Korea University, Seoul 02841, Republic of Korea}

\author{Yong-Su Kim}
\email{yong-su.kim@kist.re.kr}
\affiliation{Center for Quantum Technology, Korea Institute of Science and Technology (KIST), Seoul 02792, Republic of Korea}
\affiliation{Division of Quantum Information, KIST School, Korea University of Science and Technology, Seoul 02792, Republic of Korea}

\begin{abstract}
Reference frame mismatch among communication parties introduces errors in quantum cryptographic protocols. As the number of
participants increases, aligning reference frames becomes increasingly difficult, complicating multipartite quantum cryptographic
implementations. Here, we theoretically and experimentally investigate multipartite reference-frame-independent (RFI) quantum
cryptographic communication using Greenberger-Horne-Zeilinger (GHZ) states. We generalize the bipartite RFI security parameter $C$ to an
$N$-party parameter $C_N$ and derive the asymptotic secret key rate expressed solely in terms of experimentally accessible quantities. We analyze the key rate under global and local depolarizing noise models and find that increasing the number of parties $N$ enhances robustness against global depolarizing noise while increasing vulnerability to local channel noise. We also present a proof-of-principle experimental demonstration of four-party RFI quantum cryptographic communication using four-photon GHZ states, confirming the reference-frame invariance of both the $C_4$ parameter and the secret key rate under various reference frame rotations.
\end{abstract}

\maketitle

\section{Introduction}
Quantum key distribution (QKD) enables information-theoretically secure distribution of random bit strings between two distant parties, Alice and Bob~\cite{BB84, E91}. Significant theoretical and experimental effort has been devoted to improving the QKD performances such as security and practicality~\cite{lo2012measurement,diamanti2016practical,liao2017satellite,lucamarini2018overcoming,park2018practical,park20222}. While most of QKD protocols focus on the secret key distribution between {\it two} parties, it is intriguing to expand the protocols among multiple parties. This multipartite quantum communication offers practical cryptographic applications such as quantum cryptographic conferencing~\cite{murta2020quantum,hahn2020anonymous,proietti2021experimental} and quantum secret sharing~\cite{hillery1999quantum,tittel2001experimental,zhang2005multiparty}. 

Conventional quantum communication protocols begin with sharing a common reference frame among distant parties. Unless they use the same reference frame, secret key bits will suffer from bit-flip and/or phase-flip errors. However, in a real-world implementation, sharing a common reference frame can be cumbersome and even challenging. It becomes even more demanding in multipartite quantum communications since {\it all} the parties should have the same reference frames.

Reference-frame-independent QKD (RFI-QKD) alleviates the requirement of sharing a common reference frame and provides a way to distribute secret keys even when the reference frames are misaligned and slowly varying~\cite{laing2010reference, pramanik2017robustness, yoon2019experimental}. It is notable that the required resource of RFI-QKD is comparable to that of ordinary QKD protocols, e.g., BB84 or six-state protocols~\cite{liang2014proof,zhou2021reference}. Since its first proposal, the security and practicality have been improved~\cite{zhang2017practical,zhang2019enhanced,liu2019reference,lee2020reference,tang2022free,zhu2023improved,zhu2023reference}. However, most works focus on the secret key distribution between {\it two} parties. 

In this paper, we investigate a multiparty RFI quantum cryptographic communication protocol using Greenberger-Horne-Zeilinger (GHZ) states. We expand the security analysis of two-party RFI-QKD to the multipartite case. The main contributions of this work are threefold: (i) We generalize the reference-frame-independent security parameter $C$ to $N$-parties, defining a multiparty RFI parameter $C_N$ that remains invariant under arbitrary reference frame rotations. (ii) We derive the asymptotic secret key rate of $N$-partite RFI quantum cryptographic communication in terms of $C_N$ and the quantum bit error rate $Q_Z$, and analyze its behavior under global and local depolarizing noise models. (iii) We experimentally demonstrate the feasibility of our scheme through a proof-of-principle four-party RFI quantum cryptographic communication using four-photon GHZ states, confirming the reference-frame invariance of the secret key rate.

%We expand the security analysis of two-party RFI-QKD to the multipartite case. In particular, we generalize the reference frame rotation invariant security parameter $C$ for $N$ parties and provide the asymptotic secret key rate of $N$-partite RFI quantum cryptographic communication. We also verify the feasibility of our scheme with the proof-of-principle experiment on four-party RFI quantum cryptographic conferencing using four-photon GHZ states.

\section{Theory}

\subsection{Multiparty quantum cryptographic communication}

\begin{figure}[t]
    \includegraphics[width=3.4in]{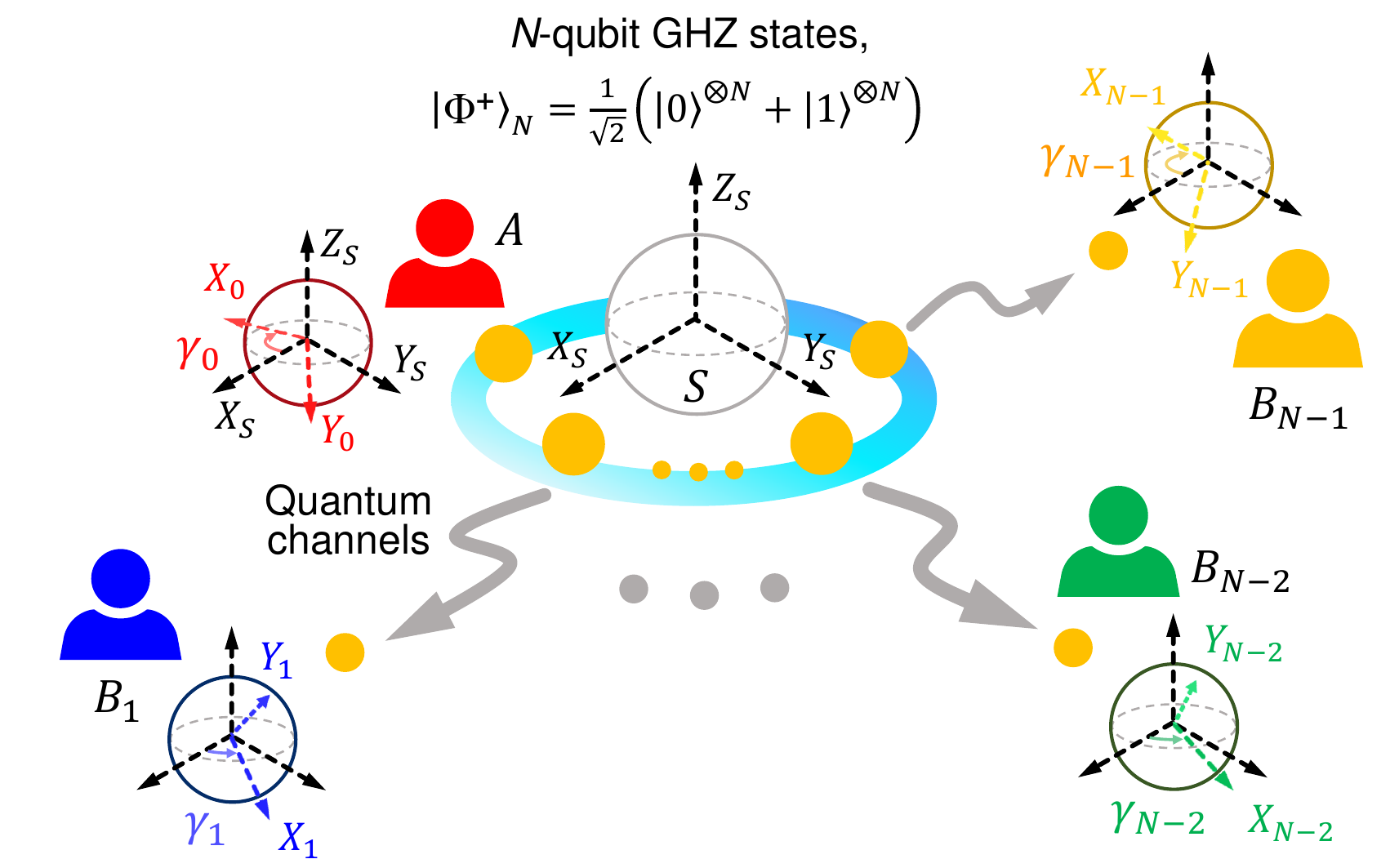}
	\caption{The schematic diagram of multiparty quantum cryptographic communication using GHZ states. Alice has $N$-partite GHZ states and distributes it into multiple Bobs through quantum channels. All communication parties, including Alice, have their reference frame misalignments as differences $\gamma_i$ between the initial frame of GHZ state preparation $S$.}
	\label{Fig:diagram_mrfi}
\end{figure}

Figure~\ref{Fig:diagram_mrfi} shows the conceptual diagram of $N$-party quantum cryptographic communication using $N$-qubit GHZ states. In the conventional multipartite quantum communication protocols~\cite{epping2017multi}, Alice ($A$), located at the central node $S$, prepares the GHZ states and distributes $N\!-\!1$ qubit states to other participants (Bobs) $B_{i}$ via quantum channels. Subsequently, Alice and all Bobs each perform projective measurements randomly chosen from the Pauli bases $X, Y$, and $Z$. Ideally, all parties share a common reference frame with identical definitions of the Pauli bases: ${X_S, Y_S, Z_S} = {X_i, Y_i, Z_i}$ for all parties $i \in {0,1,\dots,N-1}$, with $i=0$ corresponding to Alice. The participants then generate secret keys based on the correlation of measurement outcomes by estimating the information available to a potential eavesdropper (Eve).

For this entanglement-based quantum cryptographic protocol, the shared $N$-qubit GHZ state is defined as:
%%%%%%%%%%%%%%%%%%%%%%%%%%%%%%%
\begin{eqnarray}
	\vert {\rm GHZ}^{+} \rangle_{N} = \frac{1}{\sqrt{2}}(\vert 0\rangle^{\otimes N}+\vert 1 \rangle^{\otimes N}),
	\label{ghz}
\end{eqnarray}
%%%%%%%%%%%%%%%%%%%%%%%%%%%%%%%
where $|0\rangle$ and $|1\rangle$ are eigenstates of the Pauli $Z$ operator. The shared state can deviate from this ideal GHZ state due to imperfect state preparation, channel noise, decoherence, or possible information leakage to Eve. In the following analysis, we describe the noisy quantum state using the basis introduced in Ref.~\cite{epping2017multi}.

%Any attempt by Eve to extract information inevitably introduces noise into the quantum state. In the following analysis, let us describe the quantum states with the basis given as~\cite{epping2017multi}

%%%%%%%%%%%%%%%%%%%%%%%%%%%%%%%
\begin{eqnarray}
	\vert \Phi^{\pm}_{j}\rangle=\frac{1}{\sqrt{2}}(\vert0\rangle_{A}\vert j \rangle_{B} \pm \vert1\rangle_{A}\vert \bar{j}\rangle_{B}),
\end{eqnarray}
%%%%%%%%%%%%%%%%%%%%%%%%%%%%%%%
where integer indices $j$ range from 1, 2 $\cdots$,2$^{N-1}$ in a binary notation, and $\bar{j}$ denotes the binary negation of $j$. An example of this notation for four-qubit states is provided in Table~\ref{table_j}.
%%%%%%%%%%%%%%%%%%%%%%%%%%%%%%%
\begin{table}[b]
        \setlength{\tabcolsep}{0.08in}
	\begin{tabular}{@{}|c|c|c|c|c|c|c|c|c|@{}}
		\hline
		$j$ & 1 & 2 & 3 & 4 & 5 & 6 & 7 & 8 \\
		\hline
		$\vert j \rangle_{B}$ & 000 & 001 & 010 & 011 & 100 & 101 & 110 & 111 \\
		\hline
		$\vert \bar{j} \rangle_{B}$ & 111 & 110 & 101 & 100  & 011 & 010 & 001 & 000 \\
		\hline	
	\end{tabular}
        \caption{$\vert j \rangle$ and $\vert \bar{j} \rangle$ states for four-qubit states}
        \label{table_j}
\end{table}
%%%%%%%%%%%%%%%%%%%%%%%%%%%%%%%
%%%%%%%%%%%%%%%%%%%%%%%%%
%%%%%%%%%%%%%%%%%%%%%%%%%
Now, the noisy quantum state $\rho$ can be written as
%%%%%%%%%%%%%%%%%%%%%%%%%%%%%%%
\begin{eqnarray}
	\rho =\sum^{2^{N-1}}_{j=1} \lambda^{+}_{j}\vert \Phi^{+}_{j}\rangle\langle \Phi^{+}_{j}\vert + \lambda^{-}_{j}\vert \Phi^{-}_{j}\rangle\langle \Phi^{-}_{j}\vert,
	\label{Eq:noisy_GHZ}
\end{eqnarray}
%%%%%%%%%%%%%%%%%%%%%%%%%%%%%%%
with the normalization condition of $\sum^{2^{N-1}}_{j=1} \lambda^{\pm}_{j}= 1$. The initial GHZ state corresponds to $|{\rm GHZ}^+\rangle=|\Phi^+_1\rangle$ and the arbitrary noise can be represented by specific combinations of the $\lambda_j^\pm$ parameters.

In QKD protocols, this $\lambda$ is estimated from the quantum-bit error rate $Q$, given by measurement outcomes. Specifically, the bit-flip and phase-flip errors are defined as follows~\cite{scarani2009security, lee2020reference, murta2020quantum}:
\begin{align}
Q_Z &=\frac{1 - \lambda_1^+ - \lambda_1^-}{2} = \frac{1 - \langle Z^{\otimes N}\rangle}{2},\\
Q_X &=\frac{1 - \sum_{j=1}^{2^{N-1}}(\lambda_j^+ - \lambda_j^-)}{2} = \frac{1 - \langle X^{\otimes N}\rangle}{2}.
\label{Qx}
\end{align}
Note that $\langle\mathcal{O}\rangle$ is the expectation value of operator $\mathcal{O}$. {The secret key is generated from the measurement outcomes in the common $Z$ basis, while the correlations measured in the other bases are used to estimate the information potentially available to Eve.
%The bit-flip error rate can be represented with $Q$ in $Z$-basis; $Q_{Z}\!=\!1-\lambda^{+}_{1}-\lambda^{-}_{1}=\sum_{j>1} \lambda^{\pm}_{j}=(1-\langle Z^{\otimes N}\rangle)/2$, where $\langle\mathcal{O}\rangle$ is the expectation values of operator $\mathcal{O}$~\cite{lee2020reference}. The phase-flip error can be described by $X$-basis quantum bit error rate $Q_{X}=[1-\sum_{j=1}^{2^{N-1}}(\lambda^{+}_{j}-\lambda^{-}_{j})]/2=(1-\langle X^{\otimes N}\rangle )/2$~\cite{murta2020quantum}.\\

The asymptotic secret key rate $r$ is therefore determined by the balance between the correlations shared by the participants and Eve's possible information~\cite{scarani2009security,murta2020quantum}. For multiparty quantum cryptographic communication, the achievable secret key rate is limited by the weakest Alice-Bob correlation. The key rate can then be written as
%The asymptotic secret key rate $r$ for multiparty quantum cryptographic communication is generally expressed as
%%%%%%%%%%%%%%%%%%%%%%%%%%%%%%%
\begin{eqnarray}
	r=\max\lbrace \min_{i>0}[I(A\!:\!B_{i})] - I_{E},0 \rbrace,
	\label{Eq:keyrate_general}
\end{eqnarray}
%%%%%%%%%%%%%%%%%%%%%%%%%%%%%%%
where $\min_{i>0}[I(A:B_i)]$ represents the minimum mutual information shared between Alice and any Bob, and $I_E$ quantifies Eve's acquired information. The mutual information is evaluated using the binary Shannon entropy $h(x)=-x\log_{2}x - (1-x)\log_{2}(1-x)$ as
\begin{equation}
\min_{i>0}[I(A:B_i)] = 1 - h(\max[Q_{AB_i}]),
\end{equation}
with $Q_{AB_i}=\frac{1}{2}(1-\langle Z_A Z_{B_i}\rangle)$. We can estimate the bipartite quantum bit error rate $Q_{AB_i}$ for every $B_{i}$. Eve's information, represented by the Holevo bound~\cite{scarani2009security}, is computed as
\begin{equation}
I_E = S(\rho_E) - \frac{1}{2}[S(\rho_{E|0}) + S(\rho_{E|1})],
\end{equation}
where $S(\rho)$ denotes the von Neumann entropy of state $\rho$, and $S(\rho_{E|k})$ are conditional entropies given the key bit $k$. Following the standard QKD security analysis, we estimate the conditional entropies from the quantum bit error rate as $S(\rho_{E|0}) = S(\rho_{E|1}) = h(Q_Z)$. Additionally, assuming that Eve holds a purification of the participants' state, we estimate $S(\rho_E)=H(\Lambda)$, where $\Lambda=\{\lambda_j^{\pm}\}$ denotes the estimated weights of the GHZ-basis noise components and $H(\Lambda)=-\sum_{j,\pm}\lambda_j^\pm\log_2\lambda_j^\pm$ is the Shannon entropy of this distribution. Thus, estimating the noise parameters $\Lambda$ corresponds directly to quantifying Eve's information leakage.

%Additionally, we estimate that $S(\rho_E)=H(\Lambda)$, with $\Lambda=\{\lambda_j^\pm\}$ denoting the estimated GHZ-basis noise probabilities, assuming Eve holds a purification of the participants' state~\cite{scarani2009security}. Thus, estimating the noise parameters $\Lambda$ corresponds directly to quantifying Eve's information leakage.

There are various methods to calculate the error terms $\Lambda$ in quantum communication protocols, each corresponding to a distinct protocol due to the significance of the estimation technique in determining a protocol’s characteristics and performance~\cite{scarani2009security,murta2020quantum}. To lay the groundwork for addressing robustness against reference-frame misalignments in quantum communication, we begin by describing the multiparty six-state protocol, which naturally utilizes all Pauli bases (${X,Y,Z}$). In particular, utilizing all measurements offers explicit noise characterization and enhanced security checks, such as the \textit{external depolarization procedure}~\cite{epping2017multi}. This procedure partially depolarizes noisy states by selectively flipping measurement outcomes conditioned on the number of Y-basis measurements at each round, simplifying noise terms to $\lambda_{j>1}^+=\lambda_{j>1}^-$. This method resembles Pauli twirling, but unlike general twirling, external depolarization does not uniformly flatten noise terms due to its restricted conditional processing~\cite{epping2017multi}.

Hence, the secret key rate for the multiparty six-state protocol is given as
\begin{eqnarray}\nonumber
	r_{N}&=&(1-\frac{Q_{Z}}{2}-Q_{X})\log_2(1-\frac{Q_{Z}}{2}-Q_{X})\\\nonumber
	&&+~(Q_{X}-\frac{Q_{Z}}{2})\log_2 (Q_{X}-\frac{Q_{Z}}{2})\\\nonumber
	&&+~(1-Q_{Z})(1-\log_2(1-Q_{Z}))\\
	&&-~h(\max[Q_{AB_{i}}]).
	\label{Eq:keyrate_Nsix}
\end{eqnarray}
Equivalently, it can be represented in terms of $\lambda_j^\pm$ as
%%%%%%%%%%%%%%%%%%%%%%%%%%%%%%%
\begin{eqnarray}\nonumber
	%	r_{N}&=&S(E\vert K) - S(E)-\max_{i\in\lbrace 1, \cdots, N\rbrace} H(K_{i}\vert K)-H(K_{i}),\\\nonumber
	r_{N}&=&\lambda^{+}_{1}\log_{2}\lambda^{+}_{1}+\lambda^{-}_{1}\log_{2}\lambda^{-}_{1}\\\nonumber
	&+&(\lambda^{+}_{1}+\lambda^{-}_{1})(1-\log_{2}(\lambda^{+}_{1}+\lambda^{-}_{1}))\\
	&-&~h(\max[Q_{AB_{i}}]).
	%&-&h(\max_{j>1}\lambda_{j}).
	\label{Eq:keyrate_information_2}
\end{eqnarray}
%%%%%%%%%%%%%%%%%%%%%%%%%%%%%%%
%Before saying the reference-frame independent features, we start with multiparty six-state protocols, performing measurements in the three bases $\lbrace X,Y, Z \rbrace$. Using all Pauli bases leads not only to utilizing explicit noise information for security checks, but also to using the other valuable techniques available; for example, Ref.~\cite{epping2017multi} shows \textit{external depolarization procedure}, the way for reduction to depolarized states with a set of local operations. By flipping some outcomes depending on the number of Y-basis measurements at each round, any $N$-qubits state can be brought to the depolarized state having the error terms $\lambda^{+}_{j>1}=\lambda^{-}_{j>1}$, which makes the phase-flip error as $Q_{X}=(1-\lambda^{+}_{1}+\lambda^{-}_{1})/2$. This procedure is similar to Pauli twirling, which makes a flattened noise character from a set of randomizing operations~\cite{nielsen2010quantum}. However, unlike the general twirling process, \textit{external depolarization procedure} does not make a flatten between $\lambda_{j}$s because of restricted conditional post-processing, not utilizing all Pauli bases. Still, it can draw valuable features in terms of security checks, then the asymptotic secret key rate of the multiparty six-state protocol is given by

\subsection{Multiparty RFI security parameter, $C_{N}$}

Now, we consider the scenario where the reference frames of the communication parties are misaligned by certain rotation angles relative to the central node $S$, as illustrated in Fig.~\ref{Fig:diagram_mrfi}. In typical quantum communication protocols, one Pauli basis is naturally aligned—for example, right and left-handed circular polarization states in polarization-encoded free-space protocols or fast and slow pulses in time-bin encoded fiber-based protocols. Here, we choose the Pauli $Z$ basis as the invariant basis.

Each communication party, $A$ and $B_i$, has its own reference frame, defined by the measurement bases ${X_i, Y_i, Z_i}$. Relative to the central node’s reference frame ${X_S,Y_S,Z_S}$, the measurement bases can be related as~\cite{laing2010reference}
%%%%%%%%%%%%%%%%%%%%%%%%%%%%%%%
\begin{eqnarray}
	Z_S&=&Z_i,\nonumber\\
	X_S&=&\cos\gamma_i X_i+\sin\gamma_i Y_i,\label{Eq:basis_relation}\\
	Y_S&=&\cos\gamma_i Y_i-\sin \gamma_i X_i,\nonumber
\end{eqnarray}
%%%%%%%%%%%%%%%%%%%%%%%%%%%%%%%
where $\gamma_i$ denotes the rotation angle of each participant's reference frame, including Alice ($i=0$). While the $Z$ basis remains invariant under the reference frame rotation, $X$ and $Y$ bases  vary with $\gamma_i$. Thus, one can fail to generate secret keys due to increasing quantum bit error rates $Q_X$ in conventional multipartite quantum communication protocol.

RFI-QKD addresses this issue, enabling secure key generation despite misaligned reference frames~\cite{laing2010reference}. In RFI-QKD, secret keys are generated using the invariant $Z$ basis, with security monitored through a reference-frame-independent parameter $C$, derived from measurement results in the $X$ and $Y$ bases. Here, we extend this concept to multiparty quantum communication by generalizing the parameter $C$ in two ways: experimentally measured expectation values and security analysis involving error terms $\Lambda$.

Considering that the rotations are relative, the rotation errors in measurement bases can be equivalently represented as errors in the initial GHZ state, transforming it into
%%%%%%%%%%%%%%%%%%%%%%%%%%%%%%%
\begin{equation}
	\vert \Phi^{+}\rangle_{N} \rightarrow \vert \Phi(\Gamma)\rangle_N = \frac{1}{\sqrt{2}}(\vert 0\rangle^{\otimes N}+e^{i \Gamma}\vert 1 \rangle^{\otimes N}),
	\label{Eq:GHZ_gamma}
\end{equation}
%%%%%%%%%%%%%%%%%%%%%%%%%%%%%%%
where $\Gamma=\sum_{i}\gamma_{i}$. Thus, expectation values for measurement outcomes can be expressed as ${\rm Tr}(\rho_{\Phi_{\Gamma}}\mathcal{M}^{\otimes N}_{XYZ})$, where $\rho_{\Phi_{\Gamma}}=\vert \Phi(\Gamma)\rangle_N\langle \Phi(\Gamma)\vert$ and $\mathcal{M}^{\otimes N}_{XYZ}$ is the $N$-length Pauli string consisting of $\lbrace X, Y, Z\rbrace$. For example, for two qubits, $\mathcal{M}^{\otimes2}_{XYZ}\in\lbrace XX, XY, XZ, YX, YY, YZ, ZX, ZY, ZZ\rbrace$. In order to obtain the reference frame rotation invariant parameter, we utilize an expectation value of $\mathcal{M}^{\otimes N}_{XY}$ representing the Pauli bases combinations in $\lbrace X, Y\rbrace$ bases. One can find that the expectation values of $N$-length Pauli string in $X$ and $Y$ combination satisfy
\begin{eqnarray}\nonumber
	\langle Y^{\otimes n}X^{\otimes N-n} \rangle &=&{\rm tr} \left[
	\rho_{\Phi_{\Gamma}}
	\cdot
	\begin{pmatrix}
		0 & \dots & i^{n} \\
		\vdots &     & \vdots \\
		(-i)^{n} & \dots & 0 \\
	\end{pmatrix}\right]\\\nonumber
	&=&\frac{i^{n}e^{i \Gamma}+(-i)^{n}e^{-i \Gamma}}{2}\\\nonumber
	&=&\cos{\Gamma} ~~~~~~\! {\rm for} ~~~~ n=4m,\\\nonumber
	&=&-\sin{\Gamma} ~~~ {\rm for} ~~~~ n=4m+1,\\\nonumber
	&=&-\cos{\Gamma} ~~~ {\rm for} ~~~~ n=4m+2,\\
	&=&\sin{\Gamma} ~~~~~~ {\rm for} ~~~~ n=4m+3, \\\nonumber
	%&=&\cos{\frac{n \pi}{2}}\cos{\Gamma}-\sin{\frac{n \pi}{2}}\sin{\Gamma}.
	\label{Eq:general_YXX}
\end{eqnarray}
where $m\in\lbrace 0, 1, 2, \cdots \rbrace$. The permutations of the Pauli strings also give the same results.
Hence, the generalized multiparty parameter $C_N$ for RFI-QKD is defined as
%%%%%%%%%%%%%%%%%%%%%%%%%%%%%%%
\begin{eqnarray}
	C_{N}=\sum_{i=1}^{2^{N}} \langle \mathcal{M}^{\otimes N,i}_{XY} \rangle^{2},
	\label{Eq:N-Cparameter}
\end{eqnarray}
%%%%%%%%%%%%%%%%%%%%%%%%%%%%%%%
where each $\mathcal{M}^{\otimes N,i}_{XY}$ denotes an operator element of $\mathcal{M}^{\otimes N}_{XY}$. Importantly, the ideal $C_N$ parameter for $N$ parties is $C_{N} = 2^{N-1}$ for any reference frame rotation $\Gamma$ in Eq.~\eqref{Eq:GHZ_gamma}.

\subsection{Derivation of $C_{N}$ from noisy states}

To apply the RFI concept to multiparty quantum cryptographic communication protocols, let us derive the reference-frame-invariant parameter $C_N$ in terms of the GHZ-basis weights $\Lambda$. The derivation follows the approach of Ref.~\cite{laing2010reference}, which starts from a state mixture to set an invariant parameter under frame rotation. The key challenge is that the parties do not know their reference frame rotations $\gamma_i$, while in the worst-case scenario Eve does. To address this, we employ two successive symmetrization steps that preserve the experimentally accessible quantities ($Q_Z$, $Q_{AB_i}$, and $C_N$) while simplifying the state structure. The resulting state takes the form of a noisy GHZ state with reference frame rotations, from which the $\gamma_i$ dependence can be eliminated by the frame-invariance of $C_N$, yielding an explicit expression for $C_N$ solely in terms of $\Lambda$.

We begin with the first symmetrization. Since $C_N$ is invariant under the transformation $X_i \rightarrow -X_i$ and $Y_i \rightarrow -Y_i$, Eve cannot distinguish whether the state is $\rho_N$ or $Z^{\otimes N}\rho_N Z^{\otimes N}$, as both yield identical $Q_Z$, $Q_{AB_i}$, and $C_N$. Thus, we can consider the state mixture $\tilde{\rho}=\frac{1}{2}(\rho_N + Z^{\otimes N}\rho_N Z^{\otimes N})$, which can be written as
%%%%%%%%%%%%%%%%%%%%%%%%%%%%%%%
\begin{eqnarray}\nonumber
	\tilde{\rho}=\sum^{2^{N-1}}_{j=1}&&\mu^{+}_{j}\vert \Phi^{+}_{j}\rangle\langle \Phi^{+}_{j}\vert + \mu^{-}_{j}\vert \Phi^{-}_{j}\rangle\langle \Phi^{-}_{j}\vert\\
	&~&+~\frac{\phi_{j}}{2}\vert \Phi^{-}_{j}\rangle\langle \Phi^{+}_{j}\vert + H.c,\\\nonumber
\end{eqnarray}
%%%%%%%%%%%%%%%%%%%%%%%%%%%%%%%
where $\sum^{2^{N-1}}_{j=1} \vert \mu^{\pm}_{j}\vert^{2} = 1$ and $H.c$ denotes the Hermitian conjugate. Writing $\tilde{\rho}\rightarrow\tilde{\rho}(\vec{\phi})$ with $\vec{\phi}=\lbrace \phi_{1},\cdots,\phi_{2^{N-1}}\rbrace$, the $C_N$ parameter of this state is
%%%%%%%%%%%%%%%%%%%%%%%%%%%%%%%
\begin{eqnarray}
	C_{N}=2^{N-1}\sum_{j=1}^{2^{N-1}}(\mu^{+}_{j}-\mu^{-}_{j})^{2}+\text{Im}(\phi_{j})^{2}.
\end{eqnarray}
%%%%%%%%%%%%%%%%%%%%%%%%%%%%%%%

For the second symmetrization, we note that $C_N$ depends only on $\text{Im}(\phi_j)^2$, so it is unchanged under $\vec{\phi}\rightarrow -\vec{\phi}^{*}$. We thus define $\acute{\rho}(\vec{\Pi})=\frac{1}{2}(\tilde{\rho}(\vec{\phi})+\tilde{\rho}(-\vec{\phi}^{*}))$ with $\Pi_{j}=\text{Im}(\phi_{j})$. Setting $\acute{\Pi}_{j}=\sqrt{(\mu^{+}_{j}-\mu^{-}_{j})^{2}+\Pi_{j}^{2}}$, we obtain $\lambda^{\pm}_{j}=\frac{1}{2}(\mu^{+}_{j}+\mu^{-}_{j}\pm\acute{\Pi}_{j})$ and $\cos^{2}\gamma_{j}=\frac{1}{2}+(\mu^{+}_{j}-\mu^{-}_{j})/\acute{\Pi}_{j}$, so that $\acute{\rho}$ takes the form of a noisy GHZ state with reference frame rotations $\gamma_j$:
%%%%%%%%%%%%%%%%%%%%%%%%%%%%%%%
\begin{eqnarray}\nonumber
	\acute{\rho}(\vec{\Pi})=\sum_{j=1}^{2^{N-1}}\lambda^{+}_{j}\vert\Phi^{+}_{j}(\gamma_{j})\rangle\langle\Phi^{+}_{j}(\gamma_{j})\vert + \lambda^{-}_{j}\vert\Phi^{-}_{j}(\gamma_{j})\rangle\langle\Phi^{-}_{j}(\gamma_{j})\vert,\\
\end{eqnarray}
%%%%%%%%%%%%%%%%%%%%%%%%%%%%%%%
where $\vert\Phi^{\pm}_{j}(\gamma_{j})\rangle=\frac{1}{\sqrt{2}}(\vert j \rangle \pm e^{i\gamma_{j}}\vert \bar{j} \rangle)$. Since this state has the same structure as Eq.~\eqref{Eq:noisy_GHZ} and $C_N$ is frame-invariant, the $\gamma_j$ dependence drops out, and $C_N$ is expressed solely in terms of $\vec{\lambda}^{\pm}_{j}$:
%%%%%%%%%%%%%%%%%%%%%%%%%%%%%%%
\begin{eqnarray}
	C_{N}=2^{N-1}\sum_{j=1}^{2^{N-1}}(\lambda^{+}_{j}-\lambda^{-}_{j})^{2},
	\label{Eq:N-Cparameter_error}
\end{eqnarray}
%%%%%%%%%%%%%%%%%%%%%%%%%%%%%%%
demonstrating its invariance under arbitrary reference frame rotations $\gamma$. With the \textit{external depolarization procedure} $\lambda^{+}_{j>1}=\lambda^{-}_{j>1}$, Eq.~\eqref{Eq:N-Cparameter_error} simplifies to
%%%%%%%%%%%%%%%%%%%%%%%%%%%%%%%
\begin{eqnarray}
	C_{N}=2^{N-1}(\lambda^{+}_{1}-\lambda^{-}_{1})^{2},
	\label{Eq:N-Cparameter_error_depol}
\end{eqnarray}
%%%%%%%%%%%%%%%%%%%%%%%%%%%%%%%
which plays the same role as $Q_{X}$ in Eq.~\eqref{Qx}. In subsequent sections, we use $C_{N}$ to estimate Eve's information $I_{E}$.

\subsection{Multiparty RFI-QKD via noisy channels}
%%%%%%%%%%%%%%%%%%%%%%%%%
\begin{figure}[b]
	\includegraphics[width=3.4in]{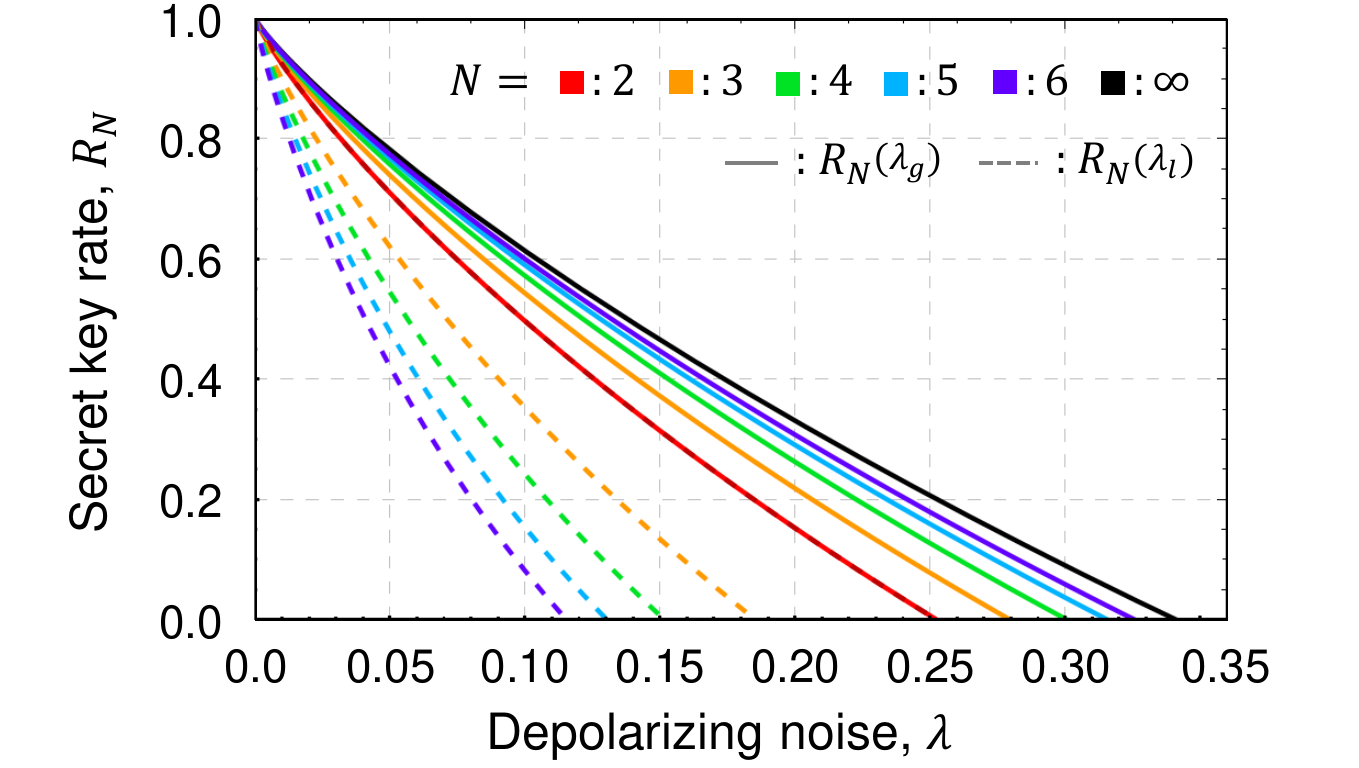} 
	\caption{The asymptotic key rates of multiparty RFI quantum cryptographic communications in terms of the \textit{global} and \textit{local} depolarizing noise of $\lambda_{g}$ and  $\lambda_{l}$, respectively. As the number of parties $N$ increasing, $R_N$ on the \textit{global} and \textit{local} depolarizing noise show different trends.}
	\label{Fig:sim_MRFI_six}
\end{figure}
%%%%%%%%%%%%%%%%%%%%%%%%%

Recall that the information leakage $I_{E}$ can be estimated using the noise parameters $\Lambda$. From previous sections, the parameters $Q_{Z}, Q_{X},$ and $C_{N}$ are expressed explicitly in terms of $\Lambda$. Notably, since the definition of $C_{N}$ involves measurements in two Pauli bases $\lbrace X, Y \rbrace$, the RFI concept naturally applies to multiparty six-state protocols without additional requirements. By substituting $\lambda^{\pm}_1$ in Eq.~\eqref{Eq:keyrate_information_2} with $C_{N}$ and $Q_{Z}$ via the relationship $\lambda^{\pm}_1=((1-Q_{Z})/2\pm\sqrt{C_{N}/2^{N+1}})$, we arrive at one of the central results of this work--the asymptotic secret key rate of the multiparty RFI-six-state quantum communication protocol expressed solely in terms of the experimentally accessible quantities $Q_Z$, $C_N$, and $Q_{AB_i}$:
%%%%%%%%%%%%%%%%%%%%%%%%%%%%%%%
\begin{eqnarray}\nonumber
	R_{N} &=& (\frac{1-Q_{Z}}{2}+\sqrt{\dfrac{C_{N}}{2^{N+1}}})\log_{2}(\frac{1-Q_{Z}}{2}+\sqrt{\dfrac{C_{N}}{2^{N+1}}}) \\\nonumber
	&+&(\frac{1-Q_{Z}}{2}-\sqrt{\dfrac{C_{N}}{2^{N+1}}})\log_{2}(\frac{1-Q_{Z}}{2}-\sqrt{\dfrac{C_{N}}{2^{N+1}}})\\\nonumber
	&+& (1-Q_{Z})(1-\log_{2}(1-Q_{Z})) - h(\max [Q_{AB_{i}}]).\\
	\label{Eq:keyrate_MRFI_2}
\end{eqnarray}
%%%%%%%%%%%%%%%%%%%%%%%%%%%%%%%
Note the notation change from $r$ to $R$ after applying the RFI concept. Equation~\eqref{Eq:keyrate_MRFI_2} constitutes a key result of this work: it provides a practical formula for evaluating the secure key rate of $N$-partite quantum cryptographic communication without requiring the knowledge of individual reference frame rotations $\gamma_i$, relying instead on the frame-invariant parameter $C_N$ that can be directly estimated from measurement statistics in the $X$ and $Y$ bases.

%Next, we examine the effects of depolarizing noise on multiparty RFI quantum cryptography. We distinguish two types of depolarizing channels: \textit{global} depolarizing channels characterized by noise $\lambda_{g}$, and \textit{local} depolarizing channels characterized by noise $\lambda_{l}$. The global depolarizing noise model corresponds to adding white noise directly to the GHZ state:

Next, we examine the effects of depolarizing noise on multiparty RFI quantum cryptography. We distinguish two types of depolarizing channels: \textit{global} depolarizing channels characterized by noise $\lambda_{g}$, and \textit{local} depolarizing channels characterized by noise $\lambda_{l}$. The \textit{global} depolarizing model describes imperfections in the initially prepared GHZ state at the central node before distribution, while the \textit{local} depolarizing model describes independent noise acting on each distributed qubit during transmission to the remote participants. The global depolarizing noise model corresponds to adding white noise directly to the GHZ state:

%%%%%%%%%%%%%%%%%%%%%%%%%%%%%%%
\begin{equation}
	\rho_N=(1-\lambda_{g})|\Phi^+\rangle_N\langle\Phi^+|_{N}+\lambda_{g}\frac{\mathbb{I}}{2^{N}},
	\label{GHZ state}
\end{equation}
%%%%%%%%%%%%%%%%%%%%%%%%%%%%%%%
where $\mathbb{I}$ is the $2^{N}$dimensional identity matrix. Note that this corresponds to the state after the \textit{external depolarization procedure}, while $\lambda_1^{+}=1-\frac{2^{N}-1}{2^{N}}\lambda_{g}$, otherwise $\lambda^{\pm}_{j}=\frac{\lambda_{g}}{2^{N}}$. Thus, the estimated parameters become 
%%%%%%%%%%%%%%%%%%%%%%%%%%%%%%%
\begin{eqnarray}
	Q_{Z}&=&\frac{2^{N-1}-1}{2^{N-1}}\lambda_{g},\\\nonumber
	Q_{X}&=&Q_{AB_{i}}=\frac{\lambda_{g}}{2},\\\nonumber
	C_{N}&=&2^{N-1}(1-\lambda_{g})^{2}.
	\label{Eq:N-Cparameter_depol_global}
\end{eqnarray}
%%%%%%%%%%%%%%%%%%%%%%%%%%%%%%%

In contrast, local depolarizing noise arises independently in each communication channel to the Bobs, modeled by the depolarizing map
%%%%%%%%%%%%%%%%%%%%%%%%%%%%%%%
\begin{equation}
	\mathcal{D}_{2}(\rho)=(1-\lambda_{l})\rho+\lambda_{l}\frac{\mathbb{I}_{2}}{2}.
	\label{local_depolarizing}
\end{equation}
%%%%%%%%%%%%%%%%%%%%%%%%%%%%%%%
The resulting noisy state is then
%%%%%%%%%%%%%%%%%%%%%%%%%%%%%%%
\begin{equation}
	\rho_N=\mathcal{D}^{\otimes(N-1)}_{2}|\Phi^+\rangle_N\langle\Phi^+|_{N}.
	\label{local_depolarized state}
\end{equation}
%%%%%%%%%%%%%%%%%%%%%%%%%%%%%%%
Here, the channel acts only on the $N-1$ distributed qubits, while Alice's qubit is kept at the central node. The corresponding parameters are then given by
%%%%%%%%%%%%%%%%%%%%%%%%%%%%%%%
\begin{eqnarray}
	Q_{AB_{i}}&=&\frac{\lambda_{l}}{2},\\\nonumber
	Q_{Z}&=&1-(1-\frac{\lambda_{l}}{2})^{N-1},\\\nonumber
	Q_{X}&=&\frac{1-(1-\lambda_{l})^{N-1}}{2},\\\nonumber
	C_{N}&=&2^{N-1}((1-\lambda_{l})^{N-1})^{2}.
	\label{Eq:N-Cparameter_depol_global}
\end{eqnarray}
%%%%%%%%%%%%%%%%%%%%%%%%%%%%%%%

Using these parameters, we analyze the asymptotic key rate $R_{N}$ with respect to the noise levels $\lambda_{g}$ and $\lambda_{l}$. It is of interest to investigate how the secret key rate depends on the number of parties $N$. In a realistic implementation, both noise sources coexist: the distributed GHZ state suffers from preparation imperfections ($\lambda_g$) followed by independent channel noise ($\lambda_l$) on each link. Here, we analyze the two models separately to isolate their distinct effects on the $N$-party scaling of the secret key rate.

Figure~\ref{Fig:sim_MRFI_six} presents the asymptotic secret key rate $R_N$ as a function of the global depolarizing noise ($\lambda_g$, solid lines) and the local depolarizing noise ($\lambda_l$, dashed lines) for various numbers of parties $N$. The two depolarizing noise models exhibit different $N$-scaling of the key rate. While the secret key rates for $N=2$ are identical, increasing the number of parties $N$ makes the protocol more robust to the \textit{global} depolarizing noise $\lambda_g$, but more vulnerable to the \textit{local} depolarizing noise $\lambda_l$.
 
%%%%%%%%%%%%%%%%%%%%%%%%%%%%%%%%%%%%%%%%%%%%%%%%
\begin{figure*}[t]
	\includegraphics[width=0.9\textwidth]{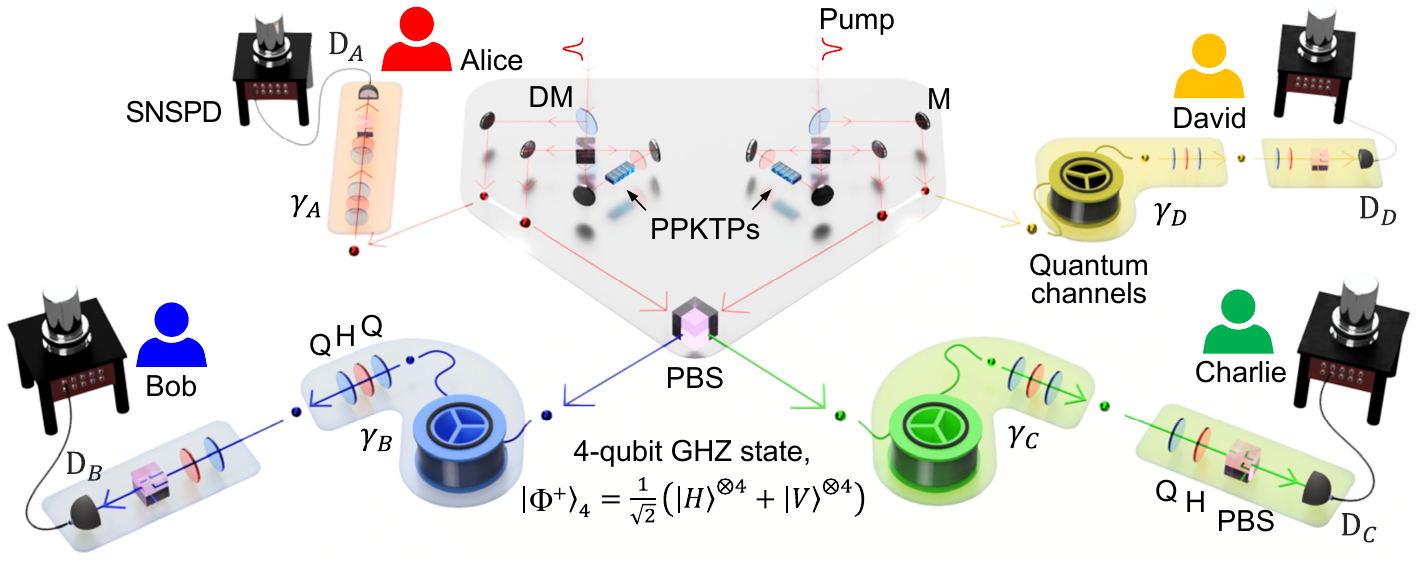}
	\caption{Experimental setups of the 4-parties RFI quantum cryptographic communication using GHZ state. M: mirror, DM: dichroic mirror, H: half waveplate, Q: quarter waveplate, PBS: polarizing beamsplitter, D: detector, SNSPD: superconducting nanowire single photon detector.}
	\label{Fig:setup_MRFI}
\end{figure*}
%%%%%%%%%%%%%%%%%%%%%%%%% 
%\textcolor{red}{The global depolarizing noise $\lambda_g$ models the imperfection of the shared GHZ state itself. In this model, the total noise $\lambda_g$ is distributed as white noise $\mathbb{I}/2^N$ across the full $2^N$-dimensional Hilbert space, so the noise per error subspace scales as $\lambda_g/2^N$, effectively diluting as $N$ increases. Combined with the external depolarization procedure ($\lambda^+_{j>1} = \lambda^-_{j>1}$)~\cite{epping2017multi}, these diluted noise terms do not contribute to the RFI security parameter $C_N$, erasing the information leakage associated with the phase-flip error contributions. This explains the robustness of $R_N$ under increasing $N$ at fixed $\lambda_g$. However, keeping $\lambda_g$ fixed while increasing $N$ is an idealization; in practice, preparing GHZ states with a larger number of parties typically incurs larger noise, so $\lambda_g$ is expected to grow with $N$.}

The different scaling behaviors can be understood from how the noise is distributed in each model. For global depolarizing model, the total noise $\lambda_g$ is distributed as white noise $\mathbb{I}/2^N$ across the full $2^N$-dimensional Hilbert space, so the noise per error subspace scales as $\lambda_g/2^N$, effectively diluting with increasing $N$. Combined with the external depolarization procedure ($\lambda^+_{j>1} = \lambda^-_{j>1}$)~\cite{epping2017multi}, these contributions do not affect the RFI security parameter $C_N$, removing the associated phase-flip information leakage. This explains the robustness of $R_N$ under increasing $N$ at fixed $\lambda_g$. However, keeping $\lambda_g$ fixed while increasing $N$ is an idealization; in practice, preparing GHZ states with a larger number of parties typically incurs larger noise, so $\lambda_g$ is expected to grow with $N$.

%The global depolarizing noise $\lambda_g$ models the imperfection of the shared GHZ state itself. The robustness of $R_N$ under increasing $N$ at fixed $\lambda_g$ originates from erasing the information leakage associated with the phase-flip error contributions, i.e., $\lambda^+_j = \lambda^-_j$ for all $j > 1$, which follows from the structure of the RFI security parameter $C_N$. This is equivalent to the effect of the external depolarization procedure in Ref.~\cite{epping2017multi}. However, keeping $\lambda_g$ fixed while increasing $N$ is an idealization; in practice, preparing higher-dimensional GHZ states typically incurs larger noise, so $\lambda_g$ is expected to grow with $N$.
 
In contrast, local depolarizing noise acts independently on each distributed qubit. Consequently, increasing $N$ increases the number of noisy channels contributing to the overall error, causing the accumulated depolarization to grow with the number of parties. This behavior explains the reduced robustness of $R_N$ against local depolarizing noise.

\section{Experimental results}\label{sec:experiment}

\subsection{Four-party RFI quantum cryptographic communication}

In order to verify the feasibility of multipartite RFI quantum cryptographic communications, we perform proof-of-principle experiments using four-photon GHZ states. 

Figure~\ref{Fig:setup_MRFI} presents the experimental setup. The four-photon GHZ states were generated at the central node $S$, which possesses two Sagnac-type Bell sources based on type-II spontaneous parametric down-conversion at 2-mm-long periodically-poled potassium titanyl phosphate (PPKTP) crystals~\cite{weston2016efficient}. The Sagnac-type Bell sources are pumped by femtosecond laser pulses centering at 780~nm, then two Bell states $\vert \phi^{+} \rangle = \frac{1}{\sqrt{2}} (\vert HV \rangle + \vert VH \rangle)$ are independently generated in polarization basis, where $\vert H \rangle$ and $\vert V \rangle$ are horizontal and vertical polarization states, respectively. By interfering two Bell states at a polarizing beamsplitter (PBS), four-photon GHZ states $\vert \Phi^+ \rangle_{4} = \frac{1}{\sqrt{2}} (\vert H \rangle^{\otimes 4} + \vert V \rangle^{\otimes 4})$ are generated when all the outputs are occupied by single-photon states with additional spectral filtering. We perform the quantum state tomography to experimentally reconstruct four-qubit quantum states, and the purity and fidelity of the reconstructed four-photon GHZ state are measured to be $P=0.81 \pm 0.02$ and $F=0.83 \pm 0.04$, respectively~\cite{james2001measurement,qi2017adaptive} Note that this non-ideal GHZ state corresponds to the noisy GHZ state of Eq.~\eqref{GHZ state} with $\lambda_{g} = 0.10 \pm 0.01$. %In practice, additional local noise ($\lambda_l$), such as dephasing and alignment drift, is present on top of the global depolarizing noise $\lambda_g$.

%We note that the quantum channels in our proof-of-principle experiment are few-meter-long optical fibers, so the contribution of local depolarizing noise $\lambda_l$ is negligibly small. Thus, the dominant noise source in our experiment is the GHZ state preparation imperfection characterized by $\lambda_g$.

Then, the central node $S$ distributes the four-photon GHZ state to four distant parties (Alice, Bob, Charlie, and David) via few-meter-long optical fibers. Note that, the central node S serves one of the communication party, Alice. In order to demonstrate the reference frame rotation, we have placed sets of half- and quarter-wave plates (HWP, QWP) after the optical fiber quantum channels. By rotating an HWP placed between two QWPs at $45^\circ$, one can implement the reference frame rotation of $\gamma_i$~\cite{yoon2019experimental}. Finally, each party $P_i$ performs projective measurements on Pauli bases $\{X_i,Y_i,Z_i\}$ by adjusting the angles of HWP and QWP in front of the PBS. The single-photon detection events are registered by superconducting nanowire single-photon detectors (SNSPD). The four-fold coincidences among all parties are obtained via a homemade coincidence counting unit (CCU)~\cite{park2015high,park2021arbitrary}.

\subsection{Estimated parameters and secure key rates}

In order to investigate the effect of reference frame rotation, we have performed the four-partite RFI quantum cryptographic communication with various sets of $\{\gamma_i\}$. We have set the communication parties as $\lbrace A,B_{1},B_{2},B_{3}\rbrace=\lbrace A,B,C,D\rbrace$ corresponding to four participants, Alice, Bob, Charlie, and David, respectively. Here, we present two cases, i.e., \textit{Case} $a$ : rotating $\gamma_{D}$ while others are fixed at $\gamma_{A}=\gamma_B=\gamma_C=0$, and \textit{Case} $b$ : rotating $\gamma_{A}$ while others are fixed at $\gamma_{B}=\gamma_C=\gamma_D=\frac{\pi}{4}$, respectively. Figure~\ref{Fig:graph_MRFI} presents the experimental results of the expectation values of the Pauli strings (the first row), quantum bit error rate in $Z$ basis $Q_Z$ and $C_4$ parameter(the second row), and the RFI secret key rates $r_{4}$ and $R_{4}$ (the third row). 

At the first row of Fig.~\ref{Fig:graph_MRFI}, some of the expectation values $\langle\mathcal{M}^{\otimes4,n}_{XY}\rangle$ curves are shown with respect to the reference frame rotation. While varying $\gamma_D$ ($\gamma_A$), expectation values of $\mathcal{M}^{\otimes4,n}_{XY}\in\lbrace XXXX,XYXX,YXXY,YXYY\rbrace$ ($\mathcal{M}^{\otimes4,n}_{XY}\in\lbrace XXXX,XXYX,YYXX,XYYY\rbrace$) show the sinusoidal interference with the average visibility of $V^{a}=0.793\pm 0.025$ ($V^{b}=0.810\pm 0.028$). The curves of $\langle XXXX \rangle$ (red circles and solid lines) correspond to the quantum bit error rate in X-basis, $Q_{X}$, in the ordinary multiparty quantum communication protocol. 

%%%%%%%%%%%%%%%%%%%%%%%%%%%%%%%%%%%%%%%%%%%%%%%%
\begin{figure}[t]
	\includegraphics[width=3.4in]{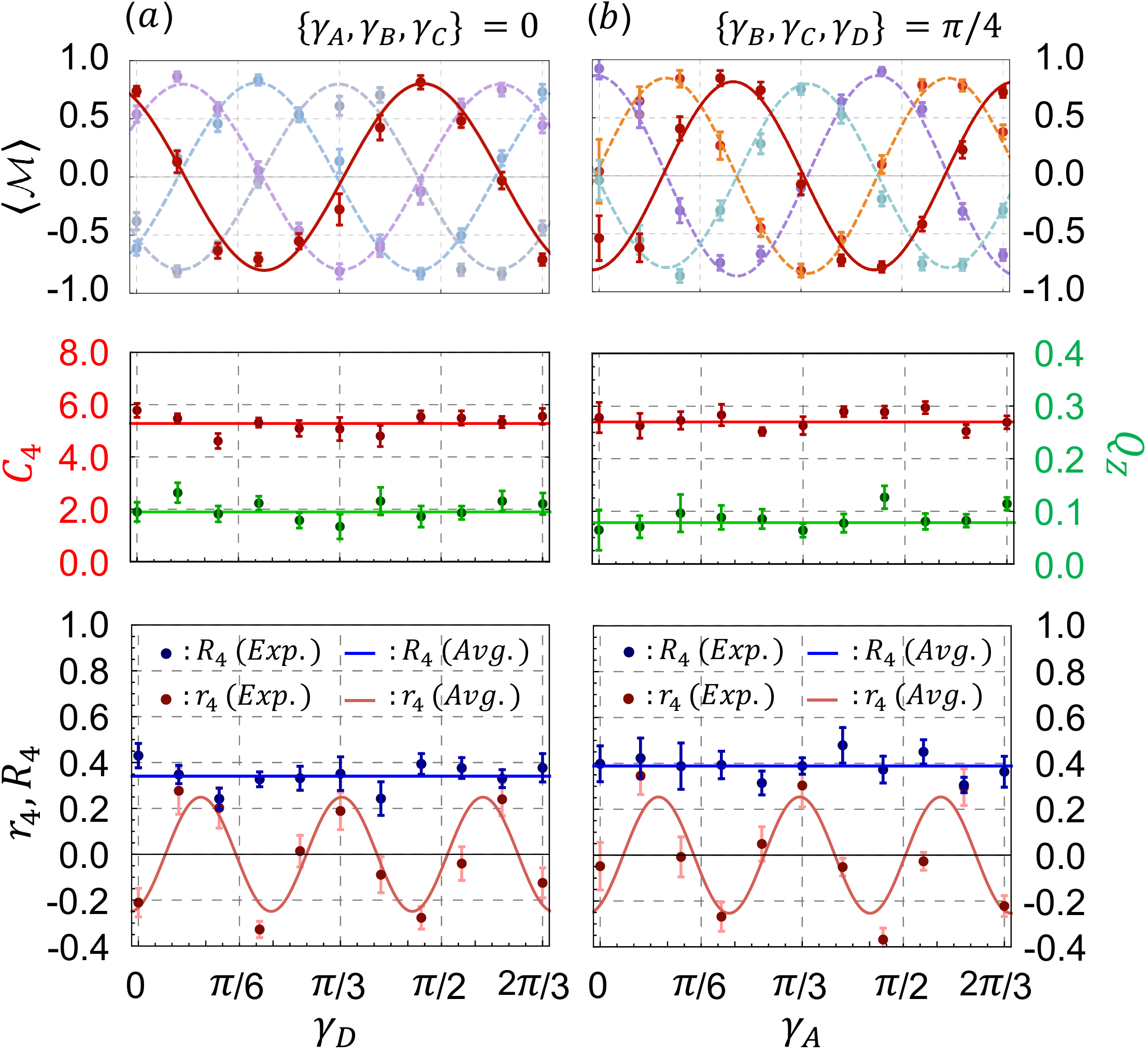}
	\caption{The experimental results of the 4-parties RFI quantum cryptographic communication using GHZ state. (a) corresponds to \textit{Case} $a$ where $\lbrace 0,0,0,\gamma_{D} \rbrace$. (b) corresponds to \textit{Case} $b$ where $\lbrace \gamma_{A},\frac{\pi}{4},\frac{\pi}{4},\frac{\pi}{4}\rbrace$. The markers correspond to the experimental data, and the error bars are derived from Monte-Carlo simulations.}
	\label{Fig:graph_MRFI}
\end{figure}
%%%%%%%%%%%%%%%%%%%%%%%%%%%%%%%%%%%%%%%%%%%%%%%%
On the other hand, the estimated quantum bit error rate in Z-basis $Q_{Z}$ (green) and $C_{4}$ parameter (red) at the second row of Fig.~\ref{Fig:graph_MRFI} are invariant under the varying reference frame. The average $Q_Z$ value is $Q^{a}_{Z}=0.098\pm 0.018$ ($Q^{b}_{Z}=0.090\pm 0.020$). The error bars are derived by Monte-Carlo simulation, which is originated from shot noise. We also found that $C^{a}_{4}=5.290\pm 0.368$ and $C^{b}_{4}=5.545\pm 0.289$, respectively. The experimentally obtained $C_4$ parameters are smaller than the theoretically expected value of $C_4 \approx 6.48$ for $\rho_4$ with the \textit{global} depolarizing noise $\lambda_g \approx 0.1$. This discrepancy is attributed to additional local noise ($\lambda_l$). When both noise sources are present, the local depolarizing channels further reduce the coherence of the distributed state, yielding $C_N = 2^{N-1}(1-\lambda_g)^2(1-\lambda_l)^{2(N-1)}$. From the experimentally measured $C_4$ values, we estimate $\lambdaƒ_l \approx 0.03$.

Finally, we have presented the estimated secret key rates $r_{4}$ and $R_{4}$ results at the third row of Fig.~\ref{Fig:graph_MRFI}. While $r_{4}$ varies with the reference frame rotation and fails to give positive key rates, $R_{4}$ clearly shows the invariant nature under the reference frame changes. In particular, the average secret key rates are \textit{Case} $a$ : $R_{4}=0.344\pm 0.061$, and \textit{Case} $b$ : $R_{4}=0.400\pm 0.050$, respectively, while both average values of $r_{4}$ give negative. This clearly demonstrates the advantage of the RFI approach.

\section{Conclusions}
In conclusion, we have proposed a multipartite RFI quantum cryptographic communication using $N$-qubit GHZ states. In particular, we have derived asymptotic secret key rates with quantum bit error rate in Z-basis, $Q_Z$, and reference frame invariant parameter, $C_N$. The feasibility of our scheme has also been explored with proof-of-principle experiments using four-photon GHZ states. We have experimentally shown that the secret keys are appropriately generated even with the reference frame rotation. Our results indicate that in real-world implementation of quantum cryptographic networks such as earth-to-satellite links~\cite{yin2017satellite} or chip-to-chip integrated photonic waveguides~\cite{zhang2014reference,sibson2017chip}, the multipartite RFI quantum cryptographic communication can provide a robust solution for better quantum secure communication performance. 

The security analysis with more practical systems, e.g., weak coherent pulses input, finite key analysis~\cite{tomamichel2012tight}, and resource-efficient implementation~\cite{liu2019reference,wang2019experimental}, would be interesting future work. We also remark that it would be promising to apply other advanced techniques developed for two-party RFI-QKD, such as measurement-device-independent (MDI) RFI-QKD~\cite{zhou2021reference, lee2020reference}, RFI-QKD with advantage distillation~\cite{zhu2023reference}, and improved RFI-QKD with the security parameter $R$~\cite{zhu2022improved}.

\section*{Acknowledgments}
\noindent This research was funded by Korea Institute of Science and Technology (26E0001), IITP-MSIT (RS-2024-00396999), NRF (RS-2024-00509800) and NIA funded by the MSIT Korea [Quantum technology test verification and consulting support in 2024].

%Korea Institute of Science and Technology (26E0001) : ORP 기고유
%IITP-MSIT (RS-2024-00396999) : QKD 과제
%NRF (RS-2024-00509800) : DV-CV 국제공동
% NIA funded by the MSIT Korea [Quantum technology test verification and consulting support in 2024]. : 테스트베드 사업

\bibliography{MRFI_ref}

@article{BB84,
  author = {Bennett, C. and Brassard, G.},
  title = {Quantum cryptography: Public key distribution and coin tossing},
  journal = {Proceedings of IEEE International Conference on Computers, Systems, and Signal Processing},
  pages = {175--179},
  year = {1984}
}

@article{E91,
  author = {Ekert, A.K.},
  title = {Quantum cryptography based on Bell’s theorem},
  journal = {Physical Review Letters},
  volume = {67},
  pages = {661},
  year = {1991}
}

@article{lo2012measurement,
  author = {Lo, H.-K. and Curty, M. and Qi, B.},
  title = {Measurement-device-independent quantum key distribution},
  journal = {Physical Review Letters},
  volume = {108},
  pages = {130503},
  year = {2012}
}

@article{diamanti2016practical,
  author = {Diamanti, E. and Lo, H.-K. and Qi, B. and Yuan, Z.},
  title = {Practical challenges in quantum key distribution},
  journal = {npj Quantum Information},
  volume = {2},
  pages = {1--12},
  year = {2016}
}

@article{liao2017satellite,
  author = {Liao, S.-K. and et al.},
  title = {Satellite-to-ground quantum key distribution},
  journal = {Nature},
  volume = {549},
  pages = {43--47},
  year = {2017}
}

@article{lucamarini2018overcoming,
  author = {Lucamarini, M. and Yuan, Z.L. and Dynes, J.F. and Shields, A.J.},
  title = {Overcoming the rate--distance limit of quantum key distribution without quantum repeaters},
  journal = {Nature},
  volume = {557},
  pages = {400--403},
  year = {2018}
}

@article{park2018practical,
  author = {Park, C.H. and et al.},
  title = {Practical plug-and-play measurement-device-independent quantum key distribution with polarization division multiplexing},
  journal = {IEEE Access},
  volume = {6},
  pages = {58587--58593},
  year = {2018}
}

@article{park20222,
  author = {Park, C.H. and et al.},
  title = {2$\times$ n twin-field quantum key distribution network configuration based on polarization, wavelength, and time division multiplexing},
  journal = {npj Quantum Information},
  volume = {8},
  pages = {1--12},
  year = {2022}
}

@article{murta2020quantum,
  author = {Murta, G. and Grasselli, F. and Kampermann, H. and Bru{\ss}, D.},
  title = {Quantum conference key agreement: A review},
  journal = {Advanced Quantum Technologies},
  volume = {3},
  pages = {2000025},
  year = {2020}
}

@article{hahn2020anonymous,
  author = {Hahn, F. and de Jong, J. and Pappa, A.},
  title = {Anonymous quantum conference key agreement},
  journal = {PRX Quantum},
  volume = {1},
  pages = {020325},
  year = {2020}
}

@article{proietti2021experimental,
  author = {Proietti, M. and et al.},
  title = {Experimental quantum conference key agreement},
  journal = {Science Advances},
  volume = {7},
  pages = {eabe0395},
  year = {2021}
}

@article{hillery1999quantum,
  author = {Hillery, M. and Bu{\v{z}}ek, V. and Berthiaume, A.},
  title = {Quantum secret sharing},
  journal = {Physical Review A},
  volume = {59},
  pages = {1829},
  year = {1999}
}

@article{tittel2001experimental,
  author = {Tittel, W. and Zbinden, H. and Gisin, N.},
  title = {Experimental demonstration of quantum secret sharing},
  journal = {Physical Review A},
  volume = {63},
  pages = {042301},
  year = {2001}
}

@article{zhang2005multiparty,
  author = {Zhang, Z.-j. and Li, Y. and Man, Z.-x.},
  title = {Multiparty quantum secret sharing},
  journal = {Physical Review A},
  volume = {71},
  pages = {044301},
  year = {2005}
}

@article{laing2010reference,
  author = {Laing, A. and Scarani, V. and Rarity, J.G. and O'Brien, J.L.},
  title = {Reference-frame-independent quantum key distribution},
  journal = {Physical Review A},
  volume = {82},
  pages = {012304},
  year = {2010}
}

@article{pramanik2017robustness,
  author = {Pramanik, T. and et al.},
  title = {Robustness of reference-frame-independent quantum key distribution against the relative motion of the reference frames},
  journal = {Physics Letters A},
  volume = {381},
  pages = {2497--2501},
  year = {2017}
}

@article{yoon2019experimental,
  author = {Yoon, J. and et al.},
  title = {Experimental comparison of various quantum key distribution protocols under reference frame rotation and fluctuation},
  journal = {Optics Communications},
  volume = {441},
  pages = {64--68},
  year = {2019}
}

@article{liang2014proof,
  author = {Liang, W.-Y. and et al.},
  title = {Proof-of-principle experiment of reference-frame-independent quantum key distribution with phase coding},
  journal = {Scientific reports},
  volume = {4},
  pages = {3617},
  year = {2014}
}

@article{zhou2021reference,
  author = {Zhou, X.-Y. and et al.},
  title = {Reference-frame-independent measurement-device-independent quantum key distribution over 200 km of optical fiber},
  journal = {Physical Review Applied},
  volume = {15},
  pages = {064016},
  year = {2021}
}

@article{zhang2017practical,
  author = {Zhang, C.-M. and Zhu, J.-R. and Wang, Q.},
  title = {Practical decoy-state reference-frame-independent measurement-device-independent quantum key distribution},
  journal = {Physical Review A},
  volume = {95},
  pages = {032309},
  year = {2017}
}

@article{zhang2019enhanced,
  author = {Zhang, H. and Zhang, C.-H. and Zhang, C.-M. and Guo, G.-C. and Wang, Q.},
  title = {Enhanced measurement-device-independent quantum key distribution in reference-frame-independent scenario},
  journal = {JOSA B},
  volume = {36},
  pages = {476--481},
  year = {2019}
}

@article{liu2019reference,
  author = {Liu, H. and Wang, J. and Ma, H. and Sun, S.},
  title = {Reference-frame-independent quantum key distribution using fewer states},
  journal = {Physical Review Applied},
  volume = {12},
  pages = {034039},
  year = {2019}
}

@article{lee2020reference,
  author = {Lee, D. and et al.},
  title = {Reference-frame-independent, measurement-device-independent quantum key distribution using fewer quantum states},
  journal = {Optics Letters},
  volume = {45},
  pages = {2624--2627},
  year = {2020}
}

@article{tang2022free,
  author = {Tang, B.-Y. and et al.},
  title = {Free-running long-distance reference-frame-independent quantum key distribution},
  journal = {npj Quantum Information},
  volume = {8},
  pages = {117},
  year = {2022}
}

@article{zhu2023improved,
  author = {Zhu, J.-R. and Wang, R. and Zhang, C.-M.},
  title = {Improved reference-frame-independent quantum key distribution: erratum},
  journal = {Optics Letters},
  volume = {48},
  pages = {468--468},
  year = {2023}
}

@article{zhu2023reference,
  author = {Zhu, J.-R. and Zhang, C.-M. and Wang, R. and Li, H.-W.},
  title = {Reference-frame-independent quantum key distribution with advantage distillation},
  journal = {Optics Letters},
  volume = {48},
  pages = {542--545},
  year = {2023}
}

@article{epping2017multi,
  author = {Epping, M. and Kampermann, H. and Bru{\ss}, D. and et al.},
  title = {Multi-partite entanglement can speed up quantum key distribution in networks},
  journal = {New Journal of Physics},
  volume = {19},
  pages = {093012},
  year = {2017}
}

@article{scarani2009security,
  author = {Scarani, V. and et al.},
  title = {The security of practical quantum key distribution},
  journal = {Reviews of modern physics},
  volume = {81},
  pages = {1301},
  year = {2009}
}

@article{yin2017satellite,
  author = {Yin, J. and et al.},
  title = {Satellite-based entanglement distribution over 1200 kilometers},
  journal = {Science},
  volume = {356},
  pages = {1140--1144},
  year = {2017}
}

@article{zhang2014reference,
  author = {Zhang, P. and et al.},
  title = {Reference-frame-independent quantum-key-distribution server with a telecom tether for an on-chip client},
  journal = {Physical Review Letters},
  volume = {112},
  pages = {130501},
  year = {2014}
}

@article{sibson2017chip,
  author = {Sibson, P. and et al.},
  title = {Chip-based quantum key distribution},
  journal = {Nature communications},
  volume = {8},
  pages = {1--6},
  year = {2017}
}

@article{tomamichel2012tight,
  author = {Tomamichel, M. and Lim, C.C.W. and Gisin, N. and Renner, R.},
  title = {Tight finite-key analysis for quantum cryptography},
  journal = {Nature communications},
  volume = {3},
  pages = {634},
  year = {2012}
}

@article{wang2019experimental,
  author = {Wang, J. and Liu, H. and Ma, H. and Sun, S.},
  title = {Experimental study of four-state reference-frame-independent quantum key distribution with source flaws},
  journal = {Physical Review A},
  volume = {99},
  pages = {032309},
  year = {2019}
}

@article{zhu2022improved,
  author = {Zhu, J.-R. and Wang, R. and Zhang, C.-M.},
  title = {Improved reference-frame-independent quantum key distribution},
  journal = {Optics Letters},
  volume = {47},
  pages = {4219--4222},
  year = {2022}
}

@article{weston2016efficient,
  author = {Weston, M.M. and et al.},
  title = {Efficient and pure femtosecond-pulse-length source of polarization-entangled photons},
  journal = {Optics express},
  volume = {24},
  pages = {10869--10879},
  year = {2016}
}

@article{james2001measurement,
  author = {James, D.F. and Kwiat, P.G. and Munro, W.J. and White, A.G.},
  title = {Measurement of qubits},
  journal = {Physical Review A},
  volume = {64},
  pages = {052312},
  year = {2001}
}

@article{qi2017adaptive,
  author = {Qi, B. and et al.},
  title = {Adaptive quantum state tomography via linear regression estimation: Theory and two-qubit experiment},
  journal = {npj Quantum Information},
  volume = {3},
  pages = {19},
  year = {2017}
}

@article{park2015high,
  author = {Park, B.K. and Kim, Y.-S. and Kwon, O. and Han, S.-W. and Moon, S.},
  title = {High-performance reconfigurable coincidence counting unit based on a field programmable gate array},
  journal = {Applied optics},
  volume = {54},
  pages = {4727--4731},
  year = {2015}
}

@article{park2021arbitrary,
  author = {Park, B.K. and Kim, Y.-S. and Cho, Y.-W. and Moon, S. and Han, S.-W.},
  title = {Arbitrary configurable 20-channel coincidence counting unit for multi-qubit quantum experiment},
  journal = {Electronics},
  volume = {10},
  pages = {569},
  year = {2021}
}

\end{document}